\pdfoutput=1
\pdfoutput=1
\pdfoutput=1
\pdfoutput=1
\pdfoutput=1
\documentclass[journal]{IEEEtran}

\ifCLASSINFOpdf
\else
\fi
\usepackage{mathtools}
\usepackage{cuted}
\usepackage{float}
\usepackage{amsmath}
\usepackage{cite}
\begin{document}

\title{Some useful approximations for  calculation of directivities of  multibeam power patterns of  large planar arrays}

\author{Javad Shabanpour*, Homayoon Oraizi, Life Senior Member, IEEE
	\newline
Department of Electrical Engineering, Iran University of Science and Technology, Narmak, Tehran 16486-13114
\newline
Email: m.javadshabanpour1372@gmail.com}

%

\maketitle

\begin{abstract}
In this paper, by adopting the superposition of terms with unequal coefficients, some useful closed-form formulas for the class of large planar arrays are presented for predicting the directivities of radiated beams. Despite the applied simplifying assumptions, the provided formulas may be used for designing a large multiple-beam planar array with a specified power pattern. Several illustrative examples are presented which are numerically demonstrate by the MATLAB software. The applicability of the proposed synthesis method verified by the achieved conformity between the simulations and theoretical predictions. Furthermore, the impact of quantization and array dimensions on the performance of power controlling as well as the limitations on the maximum scan angle are investigated. It is believed that the proposed straightforward approach for array synthesis provides a new opportunity for various applications such as multiple-target radar systems and beamforming array antenna.
\end{abstract}

\section{Introduction}
\IEEEPARstart{S}{ince} directivity is a significant figure of merit of an antenna array, several attempts have already been made to provide a convenient expression for the calculation of directivity of linear and planar arrays arrays \cite{1,2,3,4}. Multibeam antenna arrays have been widely used in communication and surveillance systems such as MIMO \cite{5}, electronic countermeasures, direct broadcasting \cite{6} and multiple-target radar systems \cite{7}. Several studies have already assisted in addressing asymmetric multibeam reflect arrays generating multiple beams with arbitrary beam directions and gain levels \cite{8,9}. Nayeri et al. \cite{10} proposed a single-feed reflectarray with asymmetric multiple beams by implementing a brute-force optimization process for producing a phase profile of reflectarray elements resulting in a high computational cost.
 Recently, by revisiting the addition theorem in the metasurface, we have introduced the concept of asymmetric spatial power divider with arbitrary power ratio levels \cite{11}. Benefiting from the semi-analytical framework and by modulating both amplitudes and phases of the meta-atoms, one can estimate the directivity ratio levels of multibeams but not their absolute values. Digital metasurfaces with switchable asymmetric multibeam are also provided in\cite{20,21,22}.
 
 In this paper, it is demonstrated that by employing a weighted combination of individual phase-only patterns in the framework of the superposition theorem, the total radiated power may be estimated  which leads to convenient closed-form formulations. It is shown that by observing some constraints, this approach is applicable with  very good approximations, which will significantly boost the speed of designing multiple beam planar arrays without resorting to any optimization procedure. Furthermore, the required number of array elements to reach the assigned directivities is also discussed. We also present a separate approach to manipulate the power pattern of multiple beams where some radiating beams have the maximum intensity towards the end-fire direction.
 As the proof of concept, several illustrative examples demonstrate the efficacy of the proposed array synthesis method. Eventually, it is observed that the full-wave simulated results have a very good agreement with our theoretical predictions. 
 \section{Results and discussion}
 Based on the superposition of the aperture fields, the additive combination of $M$ distinct constant-amplitude gradient-phase excitations yields a mixed phase-amplitude distribution, whereby both individual functionalities will appear at the same time in the superimposed array\cite{17}.
 We will demonstrate that by adding real-valued multiplicative constants, ${a_i}$, into the conventional superposition operation, one can arbitrarily manipulate the absolute directivity of each multibeam independently through a closed-form formula for a large array. In line with our outlined objective, we employ the superposition operation with unequal coefficients as follows:
\small
    \begin{equation}
{a_1}{e^{j{\phi _1}}} + {a_2}{e^{j{\phi _2}}} + ... + {a_M}{e^{j{\phi _M}}} = \,\,\left| b \right|{e^{j{\phi _T}}}
 \end{equation}
 \normalsize
where, ${e^{j{\phi _T}}}$ and $b$ carry the phase and amplitude information of a superimposed array,
 respectively.
Thanks to the fact that far-field radiation pattern and corresponding phase/amplitude excitation pattern are a Fourier transform pair\cite{12}, by taking 2D IFFT from (1), then
\small
   \begin{equation}
{a_1}{F_1}(\theta ,\varphi ) + {a_2}{F_2}(\theta ,\varphi ) + ... = {F_T}(\theta ,\varphi )
\end{equation}
 \normalsize
 where ${F_i}(\theta ,\varphi )$ represents the $i$th array factor of primary planar arrays and ${F_T}(\theta ,\varphi )$ stands for the superimposed array factor of the final planar array. Eventually, the peak directivity of a multibeam planar array toward $({\theta _i},{\varphi _i})$ can be calculated as follows\cite{18}:
 \small
     \begin{equation}
 D({\theta _i},{\varphi _i}) = \frac{{4\pi {F_T}({\theta _i},{\varphi _i})F_T^ * ({\theta _i},{\varphi _i})}}{{\int_0^{2\pi } {\int_0^{\pi /2} {{F_T}(\theta ,\varphi )F_T^ * (\theta ,\varphi )\sin \theta d\theta d\varphi } } }}
 \end{equation} 
  \normalsize
    Without loss of generality, we focus on a large planar array with two beams pointing at $({\theta _1},{\varphi _1})$ and $({\theta _2},{\varphi _2})$.
  
   In the class of large arrays in which each radiated beam has a narrow beamwidth and negligible sidelobe levels, we assume that the angular position of the maximum in the array factor for the first beam is located in the vicinity of the null of the second beam, that is, ${F_2}({\theta _1},{\varphi _1}) \simeq 0$ \cite{11}. We can also estimate the total radiated power as (E2) presented in \textbf{Table. 1}. Although the application of this simplifying assumption leads to a closed-form formalism, we will show that by implementing some constraints, this assumption is valid with a very good approximation.
   As can be seen from error values presented in \textbf{Table. 1a-c}, (E2) is valid for a planar array with the length $> 5\lambda$. But for a special case, when the emitting beams  have the same azimuth angles, the above assumption will fail (See \textbf{Table. 3d}). It should be noted that we can only apply the assumption of (E2) when we use additive combination of distinct constant-amplitude gradient-phase excitation to generate multibeams. However, other multibeam-generating methods experience significant errors. By applying the above assumptions, then: 
 \small
   	\begin{equation}
D({\theta _1},{\varphi _1}) = \frac{{4\pi a_1^2{{\left| {{F_1}({\theta _1},{\varphi _1})} \right|}^2}}}{{\int\limits_0^{2\pi } {\int\limits_0^{\pi /2} {\left[ {a_1^2{{\left| {{F_1}(\theta ,\varphi )} \right|}^2} + a_2^2{{\left| {{F_2}(\theta ,\varphi )} \right|}^2}} \right]\sin \theta d\theta d\varphi } } }}
   \end{equation}
  \normalsize
  where ${a_1}$ and ${a_2}$ are real-valued coefficients. ${F_1}$ and ${F_2}$ represent the array factor of the first and second radiated beams. In the next step, we use the Jacobian\cite{19} for applying a variable change from $d\theta d\varphi $ to $d{\psi _x}d{\psi _y}$, ($F\left( {\theta ,\varphi } \right) \to F'\left( {{\psi _x},{\psi _y}} \right)$), in which, ${\psi _x} = 2\pi \frac{d}{\lambda }(\sin \theta \cos \varphi  - \sin {\theta _{\max }}\cos {\varphi _{\max }})$ and ${\psi _y} = 2\pi \frac{d}{\lambda }(\sin \theta \sin \varphi  - \sin {\theta _{\max }}\sin {\varphi _{\max }})$. ${\theta _{\max }}$ and ${\varphi _{\max }}$ represent the angles of maximum radiation with reference to broadside direction.
  \small
  \begin{equation}
  d{\psi _x}d{\psi _y} = \left| {\begin{array}{*{20}{c}}
  	{{\raise0.7ex\hbox{${\partial {\psi _x}}$} \!\mathord{\left/
  				{\vphantom {{\partial {\psi _x}} {\partial \theta }}}\right.\kern-\nulldelimiterspace}
  			\!\lower0.7ex\hbox{${\partial \theta }$}}}&{{\raise0.7ex\hbox{${\partial {\psi _x}}$} \!\mathord{\left/
  				{\vphantom {{\partial {\psi _x}} {\partial \varphi }}}\right.\kern-\nulldelimiterspace}
  			\!\lower0.7ex\hbox{${\partial \varphi }$}}}\\
  	{{\raise0.7ex\hbox{${\partial {\psi _y}}$} \!\mathord{\left/
  				{\vphantom {{\partial {\psi _y}} {\partial \theta }}}\right.\kern-\nulldelimiterspace}
  			\!\lower0.7ex\hbox{${\partial \theta }$}}}&{{\raise0.7ex\hbox{${\partial {\psi _y}}$} \!\mathord{\left/
  				{\vphantom {{\partial {\psi _y}} {\partial \varphi }}}\right.\kern-\nulldelimiterspace}
  			\!\lower0.7ex\hbox{${\partial \varphi }$}}}
  	\end{array}} \right| = {k^2}{d^2}\sin \theta \cos \theta d\theta d\varphi 
  \end{equation}   
    \normalsize
  Since the array is large and the beam-width of each independent beam is narrow, then, the major contributions to the integral of total radiated power for the first and second beam will be in the neighborhood of ${\theta _1}$ and ${\theta _2}$ respectively. Therefore, the term $\cos \theta $ appearing in the integral of total radiated power of the first and second beams can be approximated by $\cos {\theta _1}$ and $\cos {\theta _2}$, respectively \cite{13}.
In other words, the total radiated power of a single beam due to a  large planar array  along $({\theta _i},{\varphi _i})$ direction can be written as:  
\small 
\begin{equation}
{P_{{\rm{radiation}}}}({\theta _i}) \cong \frac{1}{{\cos {\theta _i}}} \times {P_{{\rm{radiation}}}}({\rm{broadside}})
\end{equation} 
    \normalsize
Considering (5) and (6), then (4) becomes:
\small
\begin{equation}
D({\theta _1},{\varphi _1}) \cong \frac{{4\pi {k^2}{d^2}a_1^2{{\left| {{{F'}_b}(0,0)} \right|}^2}}}{{\left( {\frac{{a_1^2}}{{\cos {\theta _1}}} + \frac{{a_2^2}}{{\cos {\theta _2}}}} \right)\left( {\int {\int_\Omega  {{{\left| {{{F'}_b}({\psi _x},{\psi _y})} \right|}^2}d{\psi _x}d{\psi _y}} } } \right)}}
\end{equation}
 \normalsize
where $\Omega  = {({\psi _x})^2} + {({\psi _y})^2} \le {k^2}{d^2}$ in which $d$ represents the inter-element spacing. ${{F'}_{\rm{b}}}({\psi _x},{\psi _y})$ stands for the array factor in a broadside direction and has a uniform excitation amplitude ($\left| {{e^{j{\phi _i}}}} \right| = 1$). It could be expressed as the product of those two linear arrays, 
\small
\begin{equation}
{F'_b}({\psi _x},{\psi _y}) = {F'_b}({\psi _x}){F'_b}({\psi _y})
\end{equation}
 \normalsize
By applying (8), known as the separable or multiplication method\cite{14}, $D({\theta _1},{\varphi _1})$ becomes:
\small
\begin{equation}
\frac{{\pi a_1^2}}{{\left( {\frac{{a_1^2}}{{\cos {\theta _1}}} + \frac{{a_2^2}}{{\cos {\theta _2}}}} \right)}}\frac{{2kd{{\left| {{{F'}_b}(0)} \right|}^2}}}{{\int\limits_{ - kd}^{kd} {{{\left| {{{F'}_b}({\psi _x})} \right|}^2}d{\psi _x}} }}\frac{{2kd{{\left| {{{F'}_b}(0)} \right|}^2}}}{{\int\limits_{ - kd}^{kd} {{{\left| {{{F'}_b}({\psi _y})} \right|}^2}d{\psi _y}} }}
\end{equation}
 \begin{equation}
 D({\theta _1},{\varphi _1}) \cong \frac{{\pi a_1^2}}{{\left( {\frac{{a_1^2}}{{\cos {\theta _1}}} + \frac{{a_2^2}}{{\cos {\theta _2}}}} \right)}}{D_x}{D_y} = \frac{{\cos {\theta _1}}}{{1 + {{\left( {\frac{{{a_2}}}{{{a_1}}}} \right)}^2}\left( {\frac{{\cos {\theta _1}}}{{\cos {\theta _2}}}} \right)}}{D_{\max }}
\end{equation}
 \normalsize
Note that in deducing (10), ${D_x}$ and ${D_y}$ represent the peak directivities of linear arrays along the $x$ and $y$ directions. They are equal to ${{2L} \mathord{\left/
		{\vphantom {{2L} \lambda }} \right.
		\kern-\nulldelimiterspace} \lambda }$ for large arrays\cite{15} in which $L$ denotes the length of the array. 	${D_{\max }}$ represents the maximum directivity of a planar array equals to ${{4\pi {L^2}} \mathord{\left/
		{\vphantom {{4\pi {L^2}} {{\lambda ^2}}}} \right.
		\kern-\nulldelimiterspace} {{\lambda ^2}}}$. Following the same steps, the peak directivity of a multibeam planar array toward $({\theta _2},{\varphi _2})$ can be immediately obtained as:
	\small	
		\begin{equation}			
D({\theta _2},{\varphi _2}) = \frac{{{{\left( {\frac{{{a_2}}}{{{a_1}}}} \right)}^2}\cos {\theta _1}}}{{1 + {{\left( {\frac{{{a_2}}}{{{a_1}}}} \right)}^2}\left( {\frac{{\cos {\theta _1}}}{{\cos {\theta _2}}}} \right)}} \times {D_{\max }}
	\end{equation}
 \normalsize
  \renewcommand{\figurename}{Table}
 \renewcommand{\thefigure}{1}
  \begin{figure}[t]
 	\centering
 		\caption{Quantitative comparison between E1 and E2 when we use an additive combination of constant-amplitude gradient-phase excitation to generate multibeam. }
 	\includegraphics[width=0.4\textwidth]{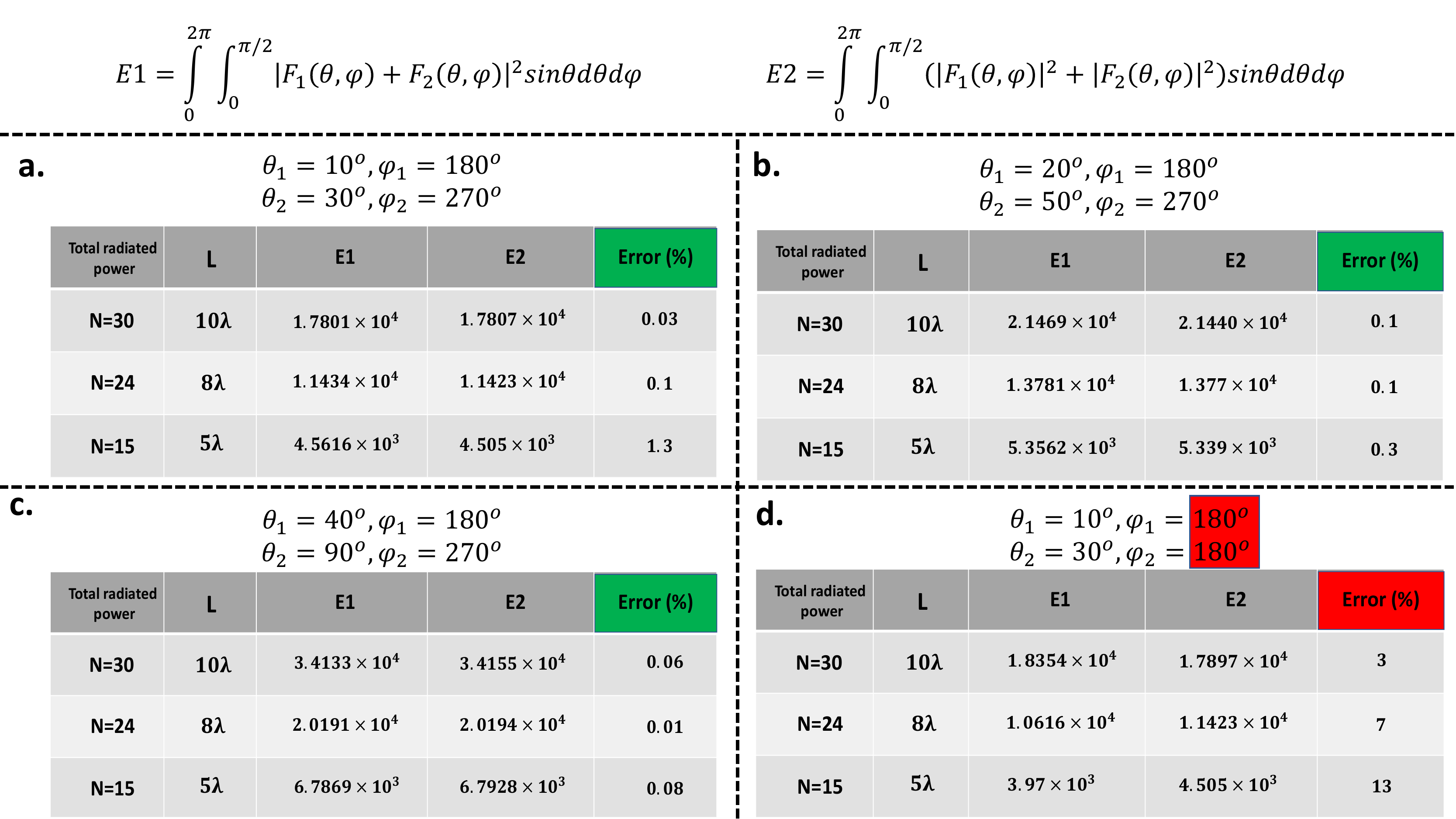}

 \end{figure} 	
For the sake of simplicity, we have defined ${D_i} = D({\theta _i},{\varphi _i})$ throughout this paper. The required number of array elements to reach the specified directivities can be obtained as:
\small	
\begin{equation}
N = \frac{\lambda }{d}\sqrt {\frac{1}{{4\pi }}\left( {\frac{{{D_1}}}{{\cos {\theta _1}}} + \frac{{{D_2}}}{{\cos {\theta _2}}}} \right)} 
 	\end{equation}
 \normalsize
For a large array with $M$ multibeams which are not scanned closer than several beam-widths to the end-fire, the generalized form of (10)-(12) can be written as below ($i,j \in [1...M]$):
\small	
\begin{equation}
({\rm{I}})\,{D_i} = \frac{{a_i^2}}{{\sum\limits_{k = 1}^M {\frac{{a_k^2}}{{\cos {\theta _k}}}} }}{D_{\max }},\,({\rm{II}})\,\frac{{{a_i}}}{{{a_j}}} = \sqrt {\frac{{{D_i}}}{{{D_j}}}} ,\,\,({\rm{III}})\,{D_{\max }} = \sum\limits_{k = 1}^M {\frac{{{D_k}}}{{\cos {\theta _k}}}}
\end{equation}
 \normalsize
Note that for a large planar array generating a single beam at the desired direction (${a_2} = 0$), it is required that all the elements have the phase gradient distribution with uniform excitation amplitude. In this case, following the previous steps (4-10), the peak directivity of a single beam planar array along direction $({\theta _1},{\varphi _1})$ is equal to $D = \pi {D_x}{D_y}\cos {\theta _1}$. This is the well-known Elliott's expression for directivity of large scanning planar arrays\cite{16}.
\newline
\textbf{End-fire beams}. Although the above formulas are useful for designing a multibeam planar array, they are valid only for radiation beams which are not scanned closer than several beam-widths to the end-fire direction. As ${\theta _i} \to {90^o}$, equation (6) is no longer valid due to the nature of the approximation. Referring to the method adopted by King and Thomas\cite{13}, after some calculations, the total radiated power for a single beam planar array at the end-fire direction can be written as:
\small	
\begin{equation}
{P_{radiation}}({\rm{endfire}}) = \frac{4}{3}\sqrt {\frac{L}{{2\lambda }}} {\mkern 1mu} {P_{radiation}}({\rm{broadside}})
\end{equation}
 \normalsize
For a two-beam planar array in which one of the radiated pencil beams points at direction $({\theta _i},{\varphi _i})$ and the other one points towards the end-fire direction, following the previous steps, the absolute directivity towards $({\theta _i},{\varphi _i})$ and the end-fire direction will be equal to (15) and (16) respectively. 
\small	
\begin{equation}
D_i = \frac{{a_1^2}}{{\left( {\frac{{a_1^2}}{{\cos {\theta _i}}} + \frac{{4a_2^2}}{3}\sqrt {\frac{L}{{2\lambda }}} } \right)}} \times {D_{\max }}
\end{equation}
\begin{equation}
{D_{endfire}} = \frac{{a_2^2}}{{\left( {\frac{{a_1^2}}{{\cos {\theta _i}}} + \frac{{4a_2^2}}{3}\sqrt {\frac{L}{{2\lambda }}} } \right)}} \times {D_{\max }}
\end{equation}
 \normalsize
For $M$ multibeams in which some radiating beams point at the end-fire direction, the above equations can be generalized. 
\newline
\textbf{limitation on the scan angle}. Equation (6) is not valid for a single beam planar array with a large scan angle. Therefore, the limitations on the maximum scan angle should be determined. By equating the two expressions for the directivity of large scanning array and directivity of an end-fire array, we have (17):
\small	
	\begin{equation}	
\frac{{4\pi {L^2}\cos {\theta _{\max }}}}{{{\lambda ^2}}} = {\mkern 1mu} 3\pi \frac{L}{\lambda }\sqrt {\frac{{2L}}{\lambda }}
\end{equation}
 \normalsize
 Therefore, eq. (6) is valid for the multiple-beam planar array with a maximum scan angle lower than that given by the limiting case as expressed below:
 \small
 	\begin{equation}
 {\theta _{\max }} \le {\mkern 1mu} {\cos ^{ - 1}}\sqrt {\frac{{9\lambda }}{{8L}}} 
\end{equation}
\normalsize
\textbf{2. Concept verification}.
The design procedures is as follows: 1. For fixed ${D_{\max }}$, by arbitrarily determining the directivity of $M$-1 beams, the directivity value of the $M$th beam is inevitably determined by (13(III)).
\newline
 2. For unknown dimensions of the array, by arbitrarily determining the directivity of $M$ beams, the length of the array ($L$) is determined by (13(III)).
 \newline
  3. For an assigned set of absolute directivities, the real-valued coefficients (${a_i}$) should be determined by  (13(II))
   \newline
4. Once the real-valued coefficients are determined, based on the superposition operation, the phase and amplitude information of the superimposed pattern ($b$ and ${\phi _T}$) can be readily obtained from (1).
 \newline
 5. Finally, the feeding networks are aimed at realizing the required local phase/amplitude excitations dictated by the superimposed pattern.

For simplicity, we focus our design on the cases of planar arrays with two and three asymmetric multiple beams. In the following, we will present four different illustrative examples in which large planar arrays are discretized in a realizable manner ($N=30$ and ${d_x} = {d_y} = {\lambda  \mathord{\left/
		{\vphantom {\lambda  3}} \right.
		\kern-\nulldelimiterspace} 3}$ leading to ${D_{\max }}=31$ dBi) with no phase/amplitude quantization.
We intend to design a two-beam generating planar array pointing at $({10^o},{180^o})$,$({30^o},{270^o})$ whose absolute directivity toward $({10^o},{180^o})$ is ${D_1}=29$ dBi for the first example (\textbf{Fig. 1a}); and the other two-beam generating planar array pointing at $({28^o},{180^o})$,$({67^o},{270^o})$ whose absolute directivity toward $({67^o},{270^o})$ is ${D_2}=25.74$ dBi for the second example (\textbf{Fig. 1b}). 
Therefore, based on (eq. 13(III)), we obtain ${D_2}=25.9$ and ${D_1}=24.41$ for the first and second examples respectively. 
By applying (eq. 13(II)), the real-valued coefficients will be equal to ${a_1}=$ 1, ${a_2=}$ 0.7 and ${a_1}=$ 1, ${a_2=}$ 1.13 for the first and second examples respectively. Quantitative comparisons between the numerical simulations
and theoretical predictions show excellent conformity (See \textbf{Fig. 1e}).
 \renewcommand{\figurename}{Figure}
  \begin{figure}[t]
	\centering
	\includegraphics[width=0.4\textwidth]{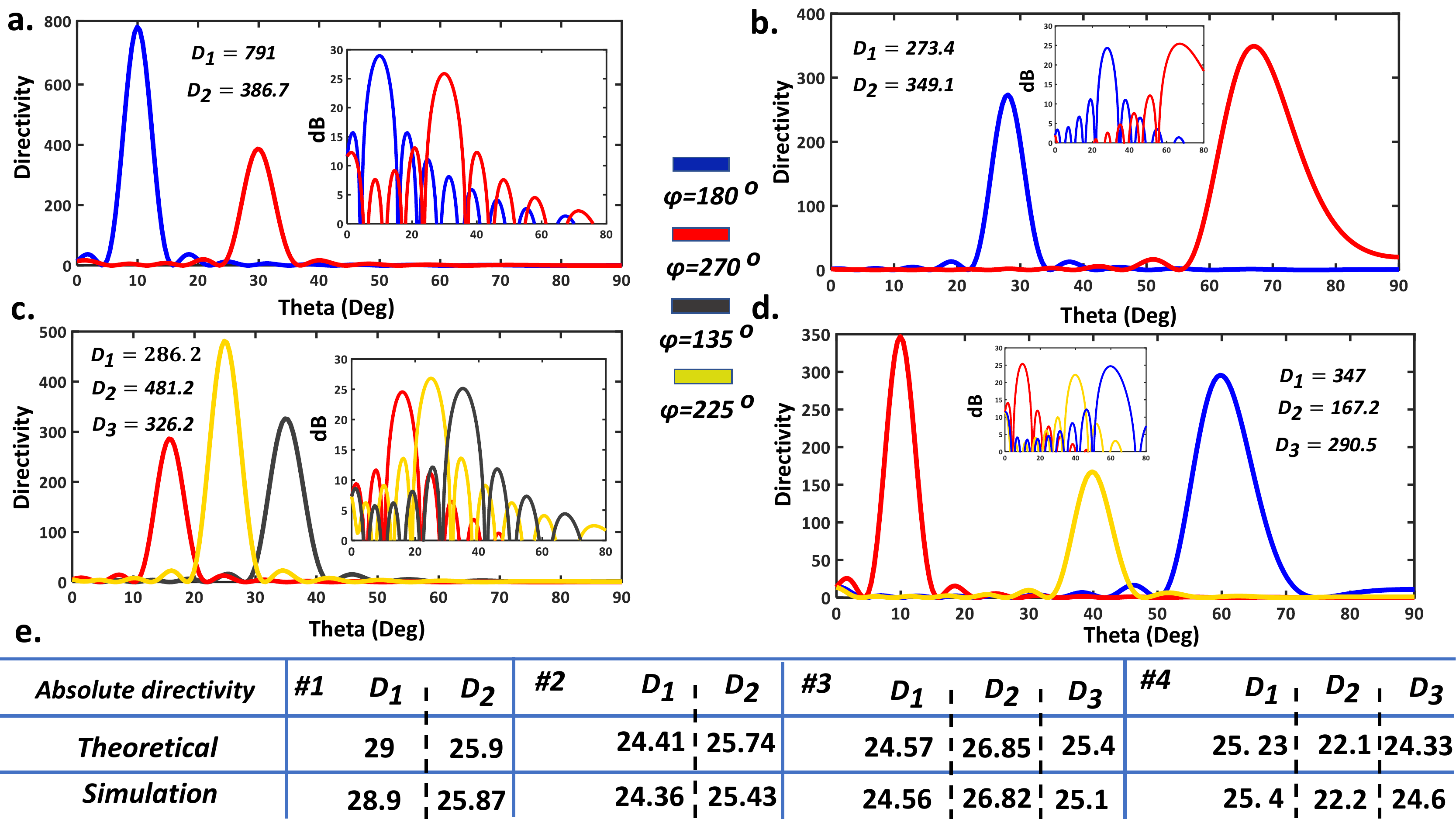}
	\caption{(a)-(d) Directivity intensity pattern (in both linear and decibel formats). (e) Comparison between simulations and theoretical predictions.}
\end{figure} 
Note that in the second example, the second beam has a large scan angle and must be checked whether it has exceeded the limit. Since the dimension of the planar array in the proposed example is $10\lambda  \times 10\lambda$  , according to (18), the limit will be equal to ${71^\circ}$  which is higher than the scan angle of the proposed beam. For the third example (See \textbf{Fig. 1c}), consider a planar array to generate three asymmetrical pencil beams towards  $({16^o},{270^o})$,$({25^o},{225^o})$ and $({35^o},{135^o})$ with ${D_1}=$  24.57 and ${D_3}=$  25.4 dBi, respectively. By applying (eq. 13(III)), the absolute directivity of the second beam is obtained as ${D_2}=$  26.85 dBi. Then, the real-valued coefficients are equal to ${a_1}=$ 1, ${a_2=}$ 1.3, ${a_3=}$ 1.1. 
After simulation, the proposed planar array generate three
beams with (${D_1}=$  24.56 and ${D_2}=$  26.82 and ${D_3}=$  25.1 dBi) that are very close to our theoretical predictions.
For the fourth example, consider an unknown number of array elements for the design of a planar array to generate three independent asymmetric beams along the directions $({10^o},{270^o})$, $({40^o},{225^o})$, $({60^o},{180^o})$ with assigned directivities of ${D_1}=$  25.23, ${D_2}=$  22.1 and ${D_3}=$  24.33. Referring to the presented formulations, the planar array should be endowed by the superimposed phase-amplitude pattern obtained by assuming (${a_1}=$ 1, ${a_2}=$ 0.7 and ${a_3}=$ 0.9) and $N=$ 28 to expose three asymmetric oriented beams with the assigned  directivities. As depicted in \textbf{Fig. 1d,e},  the simulation results are very close to our theoretical predictions.
\newline
\textbf{2.1 Quantization impact}. Up to now, we have studied continuous phase/amplitude excitations, an optimistic hypothesis that cannot be realized
in practice. We will demonstrate how quantization aggressively deteriorates the performance of the presented closed-form formulas. According to \textbf{Fig. 2}, the numerical simulations have been accomplished where both the phase and amplitude profiles describing the superimposed planar array are quantized into two (16 levels) or three bits (64 levels). Note that although the architectures with two-bit quantization fail to achieve satisfactory results in comparison to those of continuously modulated designs, the planar array with three-bit quantization (64 distinct phase/amplitude response)  operate efficiently.  
\renewcommand{\thefigure}{2}
\begin{figure}[t]
	\centering
	\includegraphics[width=0.4\textwidth]{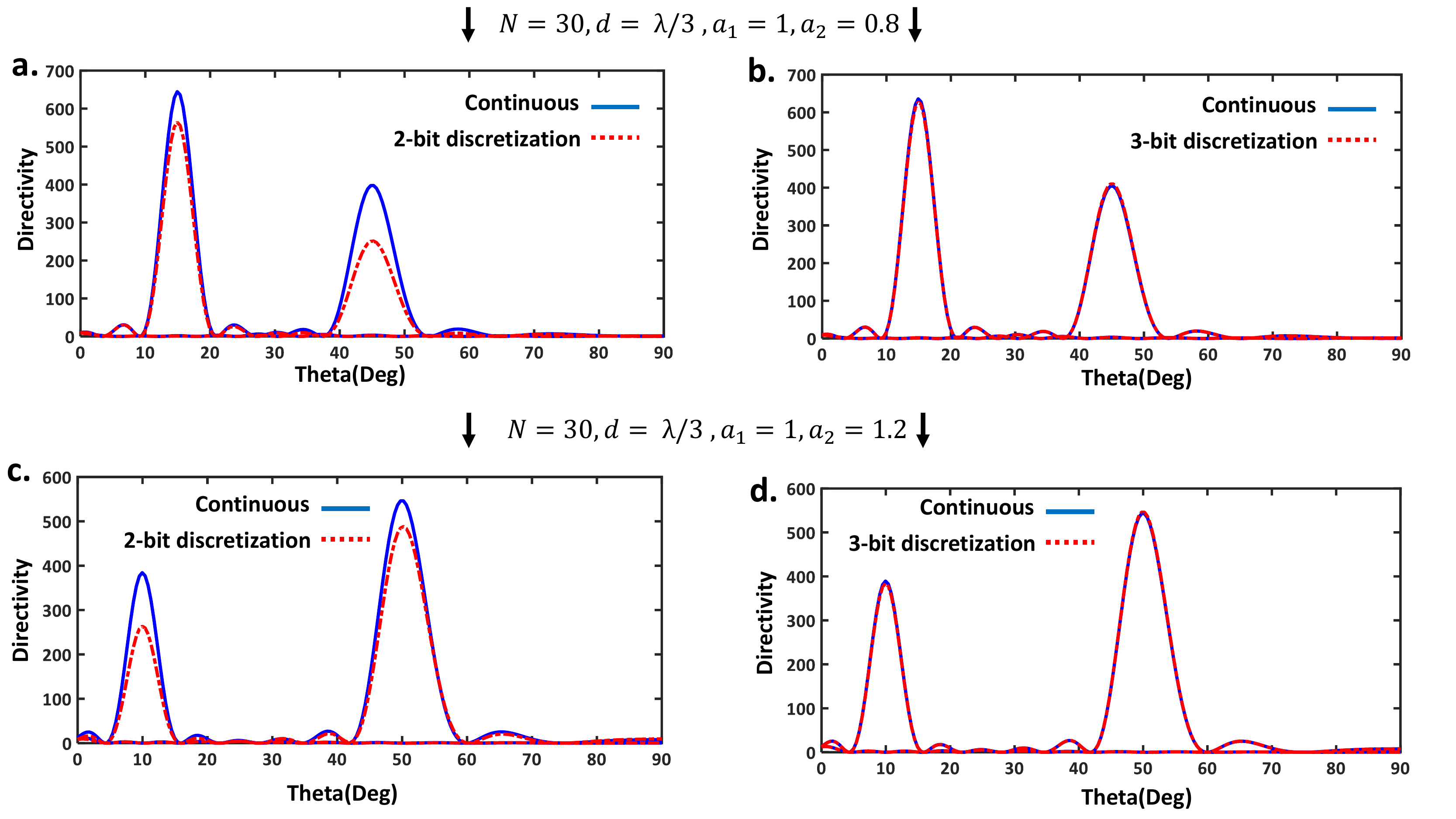}
	\caption{1D directivity intensity pattern (linear format) for a large planar array with asymmetric beams toward $({15^{\circ}},{180^{\circ}})$ and $({45^{\circ}},{270^{\circ}})$ directions assuming (a) two-bit and (b) three-bit quantization. (c,d) 1D directivity intensity pattern (linear format) for a large planar array with asymmetric beams toward $({10^{\circ}},{180^{\circ}})$ and $({50^{\circ}},{270^{\circ}})$ directions assuming two-bit and three-bit quantization.}
\end{figure} 

In the next two examples, we involve the 3-bit quantization effects in our study. For the first representation, a two-beam large planar array towards directions  $({45^{\circ}},{45^{\circ}})$ and $({65^{\circ}},{135^{\circ}})$ is designed, in which the absolute directivity of the first and second beams are ${D_1}=$  22.68 and ${D_2}=$  27.55 dBi, respectively. Based on (12), the number of elements is equal to $N=34$.
Eventually, the real-valued coefficients will be equal to ${a_1}=$ 0.57, ${a_2}=$ 1. The simulated results show that the absolute directivity values reach ${D_1}=$  23 and ${D_2}=$  27.41 dBi which have good conformity with our analytical predictions (See \textbf{Fig. 3e}). The very negligible discrepancies are attributable to the nature of approximations and quantization effects. The excitation phase and amplitude maps of the planar array are also depicted in \textbf{Fig. 3b}. The realization of such a phase-amplitude excitation is simply feasible and may be fabricated by the present technologies.

To further verify the proposed concept, the last example is dedicated to a three-beam large planar array with one beam pointing at the end-fire direction (${\theta _3} = 90^\circ$). The planar array is to generate three independent/asymmetric pencil beams along directions $({10^{\circ}},{270^{\circ}})$, $({40^{\circ}},{225^{\circ}})$ and $({90^{\circ}},{180^{\circ}})$ with assigned directivities of ${D_1}=$  22.71, ${D_2}=$  22.71 and ${D_2}=22.08$, respectively. Referring to (14) and after some mathematical calculations, the number of elements to reach the proposed directivities is $N=38$. Again, the real-valued coefficients are determined by (13(II)) as ${a_1}=$ 1, ${a_2}=$ 1 and ${a_3}=$ 0.93. Finally, the phase-amplitude pattern ($b$ and ${\phi _T}$) of superimposed planar array is obtained by (1) after quantization, it functions as a large planar array architecture that elaborately splits the input energy into three asymmetric beams with directivity values of  ${D_1}=22.64$, ${D_2}=22.57$ and ${D_3}=21.96$ dBi (See \textbf{Fig. 3c}). Observe in \textbf{Fig. 3e} that the absolute directivities of three emitting beams satisfactorily approach the assigned values with the desired tilt angles. The excitation phase and amplitude maps of such a planar array are also depicted in \textbf{Fig. 3d}. Once again, the realization of such a phase-amplitude excitation is simply feasible and may be fabricated by the current technologies.
\renewcommand{\thefigure}{3}
  \begin{figure}[t]
	\centering
	\includegraphics[width=0.4\textwidth]{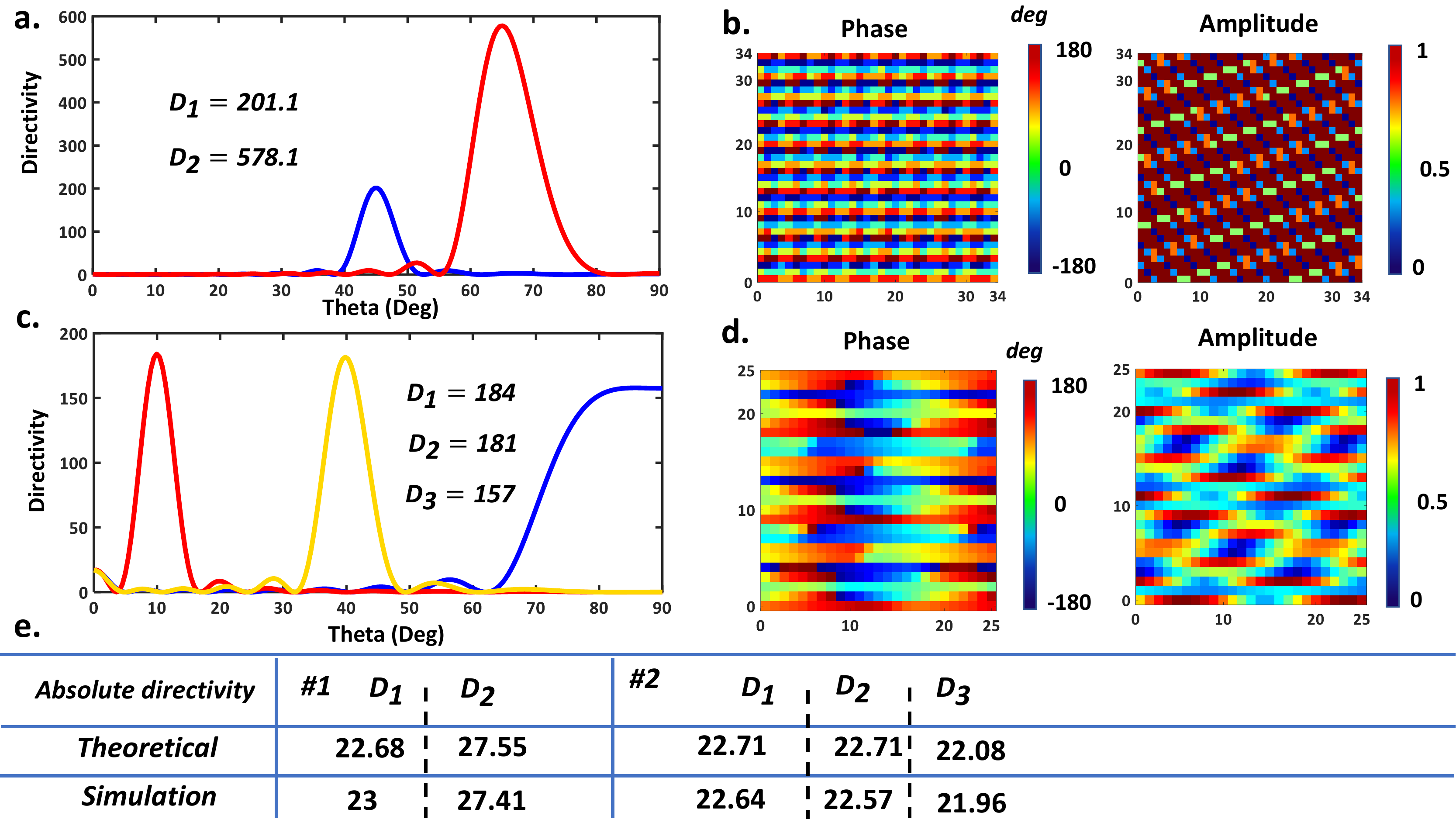}
	\caption{(a),(c) Directivity intensity pattern (in both linear and decibel formats) in which 3-bit quantization is used. (b),(d) 2D map of excitation phase and amplitude (e) Comparison between simulations and theoretical predictions.}
\end{figure} 
\newline
\newline
\textbf{2.2. Impact of the array length}. In the validation of (6), as stated before,  the beamwidth of emitting beams must be narrow. To further clarification, several numerical simulations are performed, in which planar arrays expose two differently oriented beams pointing at  directions $({10^{\circ}},{180^{\circ}})$ and $({50^{\circ}},{270^{\circ}})$ with ${a_1} = 1.2$ and ${a_2} = 1$ . Notice in \textbf{Fig. 4} that for $A < 5\lambda$,
the absolute directivities of the radiated beams do not further match with our theoretical predictions, thereby, the significant function of the planar array length in validating (6) is highlighted. Furthermore, by decreasing the length of the planar array, the limitation for the maximum scan angle which is introduced in (18) will be more restricted.
\renewcommand{\thefigure}{4}
  \begin{figure}[H]
	\centering
	\includegraphics[width=0.4\textwidth]{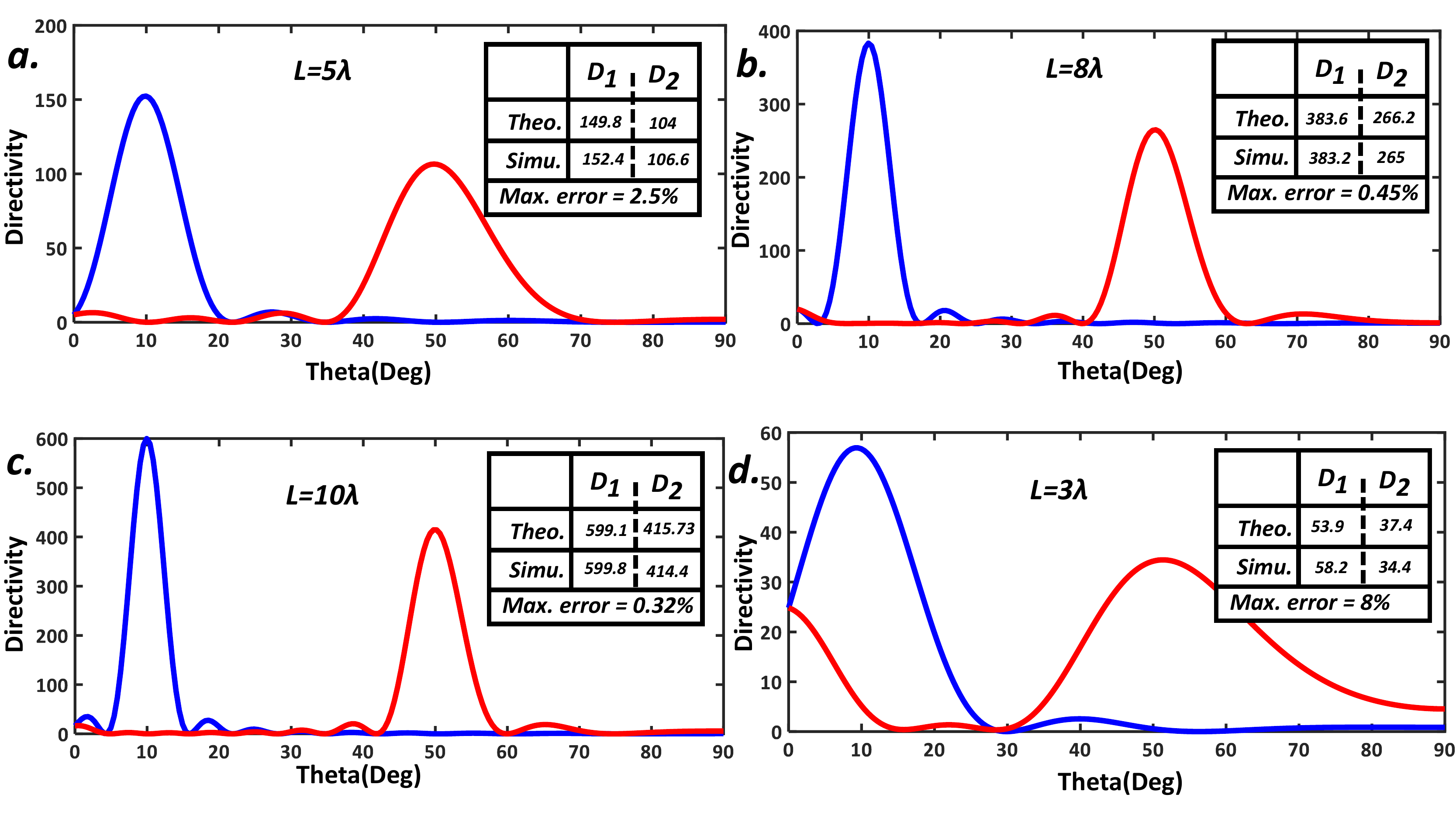}
	\caption{1D directivity intensity pattern (linear format) for planar array with asymmetric beams toward $({10^{\circ}},{180^{\circ}})$ and $({50^{\circ}},{270^{\circ}})$ directions when the length of the array is equal to (a) $5\lambda$. (b) $8\lambda$. (c) $10\lambda$. (d) $3\lambda$.}
\end{figure} 
\section{Conclusion}
In this paper, it is demonstrated that the application of additive combination of distinct constant-amplitude gradient-phase excitations for the generation of multi-beams from large planar arrays, leads to simple closed-form formulas for the calculation of total radiated powers. As the proof of concept of the proposed method, several illustrative examples are presented for the effective  prediction of the exact values of assigned directivities of multiple beams of  planar arrays at desired directions. Outstandingly, the analytical predictions estimate well the beam directivities and scanning angles. Very negligible discrepancies can be attributed to the nature of approximations and quantization effects which are interestingly less than 0.3 dB. By observing the introduced limits in the manuscript, the proposed straightforward method can be used as an alternative and  fast approach for the  design of large planar arrays generating multiple pencil-beams.

\end{document}